\documentclass[12pt]{article}
\usepackage{pdproc} 
\usepackage{epsfig}
 \def\lsi{\raise0.3ex\hbox{$<$\kern-0.75em\raise-1.1ex\hbox{$\sim$}}}
\def\gsi{\raise0.3ex\hbox{$>$\kern-0.75em\raise-1.1ex\hbox{$\sim$}}}
\newcommand{\lsim}{\mathop{\lsi}}

\newcommand{\eV}{\rm{eV}}
\newcommand{\keV}{\rm{keV}}

\def\be{\begin{equation}}
\def\ee{\end{equation}}
\def\ba{\begin{eqnarray}}
\def\ea{\end{eqnarray}}

\begin{document}

\renewcommand{\thefootnote}{\alph{footnote}}
  
\title{HOW TO FIND STERILE NEUTRINOS?\footnote{
Talk given at 12th International Workshop on Neutrinos Telescopes: Twenty Years after the Supernova 1987A 
Neutrino Bursts Discovery, Venice, Italy, 6-9 Mar 2007. }}

\author{MIKHAIL SHAPOSHNIKOV}

\address{Institut de Th\'eorie des Ph\'enom\`enes Physiques\\
Ecole Polytechnique F\'ed\'erale de Lausanne\\
CH-1015 Lausanne, Switzerland\\
 {\rm E-mail: Mikhail.Shaposhnikov@epfl.ch}}

\abstract{We describe an extention of the Standard Model (the
$\nu$MSM) by three light singlet Majorana fermions -- sterile
neutrinos, which allows to address simultaneously the problem of
neutrino oscillations and the problems of dark matter and baryon
asymmetry of the Universe.  We discuss the ways these new particles
can be searched for in astrophysical, laboratory, and accelerator
experiments.}

\normalsize\baselineskip=15pt

\section{Introduction}  
In a search for physics beyond the Standard Model (SM) one
can use different types of guidelines. A possible strategy is to
attempt to explain the phenomena that cannot be fit to the SM by
minimal means, that is by introducing the smallest possible number of
new particles without adding any new physical principles (such as
sypersymmetry or extra dimensions)  or new energy scales (such as
Grand Unified scale). An example of such a theory is the
renormalizable extension of the SM, the $\nu$MSM (neutrino Minimal
Standard Model) \cite{Asaka:2005an,Asaka:2005pn}, where three {\em
light} singlet right-handed fermions (we will be using also the names
neutral fermions, or sterile neutrinos, interchangeably)  are
introduced. The leptonic sector of the theory has the same structure
as the quark sector, i.e. every left-handed fermion has its
right-handed counterpart. This model is consistent with the data on
neutrino oscillations, provides a candidate for dark matter (DM)
particle -- the lightest singlet fermion (sterile neutrino), and can
explain the baryon asymmetry of the Universe \cite{Asaka:2005pn}. A
further extension of this model by a light singlet scalar field
allows to have inflation in the early Universe~\cite{Shaposhnikov:2006xi}.

A crucial feature of this theory is the relatively small mass scale
of the new neutral leptonic states $\sim~{\cal O}(1)~{\rm keV}-{\cal
O}(1)~{\rm GeV}$, which opens a possibility for a  direct search of
these particles. In this talk we will discuss the structure of the
model, the physical applications of the $\nu$MSM, and different
strategies to search for dark matter sterile neutrino in the Universe
and in laboratory. The accelerator experiments that can search for
two  extra singlet fermions  necessary for baryogenesis will be
discussed as well.

\section{The $\nu$MSM and its consequences}
If three singlet right-handed fermions $N_I$ are added to the
Standard Model, the most general renormalizable Lagrangian describing
all possible interactions has the form:
\be
L_{\nu MSM}=L_{SM}+
\bar N_I i \partial_\mu \gamma^\mu N_I
  - F_{\alpha I} \,  \bar L_\alpha N_I \tilde \Phi
  - \frac{M_I}{2} \; \bar {N_I^c} N_I + h.c.,
  \label{lagr}
  \ee
where $L_{SM}$ is the Lagrangian of the SM,
$\tilde\Phi_i=\epsilon_{ij}\Phi^*_j$ and $L_\alpha$
($\alpha=e,\mu,\tau$) are the Higgs and lepton doublets,
respectively, and both Dirac ($M^D = f^\nu \langle \Phi \rangle$) and
Majorana ($M_I$) masses for neutrinos are introduced. In comparison
with the SM, the $\nu$MSM contains 18 new parameters: 3 Majorana
masses of new neutral fermions $N_I$,  and 15 new Yukawa couplings in
the leptonic sector (corresponding to 3 Dirac neutrino masses, 6
mixing angles and 6 CP-violating phases).

Of course, this Lagrangian is not new and is usually used for the
explanation of the small values of neutrino masses via the see-saw
mechanism \cite{Seesaw}. The see-saw scenario assumes that the Yukawa
coupling constants of the singlet fermions are of the order of the
similar couplings of the charged leptons or quarks and that the
Majorana masses of singlet fermions are of the order of the Grand
Unified scale. The theory with this choice of parameters can also
explain the baryon asymmetry of the Universe but does not give a
candidate for a dark matter particle. Another suggestion is to fix
the Majorana masses of sterile neutrinos in $1-10$ eV energy range
(eV see-saw) \cite{deGouvea:2005er} to accommodate the LSND anomaly
\cite{Aguilar:2001ty}. This type of theory, however, cannot explain
dark matter and baryon asymmetry of the universe. The  $\nu$MSM
paradigm is to determine the Lagrangian parameters  from solid
available observations, i.e. from  requirement that it should explain
neutrino oscillations, dark matter and baryon asymmetry of the
universe in a unified way. This leads to the singlet fermion Majorana
masses  {\em smaller} than the electroweak scale, in the contrast
with the see-saw choice of \cite{Seesaw}, but much larger than few
eV, as in the eV see-saw of \cite{deGouvea:2005er}.

Let us review shortly the physical applications of the $\nu$MSM. 

{\em Neutrino masses and oscillations.}   The new parameters of the
$\nu$MSM can describe any pattern (and in particular the observed
one) of masses and mixings of active neutrinos, which is
characterized by 9 parameters only (3 active neutrino masses, 3
mixing angles, and 3 CP-violating phases). Inspite of this freedom,
the {\em absolute} scale of active neutrino masses can be established
in the $\nu$MSM from cosmology and astrophysics of dark matter
particles
\cite{Asaka:2005an,Boyarsky:2006jm,Boyarsky:2006fg,Asaka:2006rw,Asaka:2006nq}:
one of the active neutrinos must have a mass smaller than ${\cal
O}(10^{-5})$ eV. The choice of the small mass scale for singlet
fermions leads to the small values of the Yukawa coupling constants,
on the level $10^{-6}-10^{-12}$, which is crucial for explanation of
dark matter and baryon asymmetry of the Universe. 

\begin{figure}[h]
\epsfysize=4cm
\centerline{\epsffile{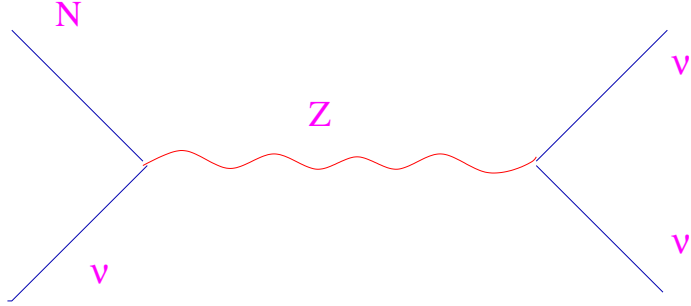}}
\caption{Main mode of dark matter sterile neutrino decay.}
\label{fig:main}
\end{figure}

{\em Dark matter.} Though the $\nu$MSM does not have any extra stable
particle in comparison with the SM, the lightest singlet fermion,
$N_1$,  may have a life-time $\tau_{N_1}$ greatly exceeding the age
of the Universe and thus play a role of a dark matter particle
\cite{Dodelson:1993je,Shi:1998km,Dolgov:2000ew,Abazajian:2001nj}. The
main decay mode is $N_1 \rightarrow 2\nu\bar\nu,2\bar\nu \nu$ (it
goes through the exchange of $Z$-vector boson, see Fig.
\ref{fig:main}) and
\be
  \tau_{N_1} = 5 \times 10^{26}\,\mbox{sec}
  \left( \frac{1~\mbox{keV}}{M_1} \right)^{5}
  \left( \frac{10^{-8}}{\theta^2} \right) ~,
\ee
where the mixing angle $\theta$ is the ratio of Dirac and Majorana
masses,
\be
\theta = \frac{m_D}{M_1}~.
\ee
For example, choosing $M_1$ in keV region and $\theta^2 \sim 10^{-8}$
leads to a life-time exceeding the age of the Universe by ten orders
of magnitude.

DM sterile neutrinos can be produced in the early Universe  via
active-sterile neutrino transition \cite{Dodelson:1993je} (probably,
this mechanism is ruled out~\cite{Asaka:2006nq}: the required Yukawa
coupling is too large to be consistent with X-ray and Lyman-$\alpha$
constraints, discussed in Section 3); via resonant active-sterile
neutrino oscillations in the presence of lepton asymmetries
\cite{Shi:1998km}; or in the inflaton (or any neutral scalar) decays
\cite{Shaposhnikov:2006xi}. DM sterile neutrino may also have other
interesting astrophysical applications~\cite{astro}.

{\em Baryon asymmetry of the Universe.} The  baryon (B) and lepton
(L)  numbers are not conserved in the $\nu$MSM. The lepton number is
violated by the Majorana neutrino masses, while  $B+L$ is broken by
the electroweak anomaly. As a result, the sphaleron processes with
baryon number non-conservation \cite{Kuzmin:1985mm}  are in thermal
equilibrium  for  temperatures $100$ GeV $ < T < 10^{12}$ GeV. As for
CP-breaking, the $\nu$MSM contains  $6$ CP-violating phases in the
lepton sector and a Kobayashi-Maskawa phase in the quark sector. This
makes two of the Sakharov conditions \cite{Sakharov:1967dj} for
baryogenesis satisfied. Similarly to the SM, this theory does not
have an electroweak phase transition with allowed values for the
Higgs mass \cite{Kajantie:1996mn}, making impossible the electroweak
baryogenesis, associated with the non-equilibrium bubble expansion.
However, the $\nu$MSM contains extra degrees of freedom - sterile
neutrinos - which may be out of thermal equilibrium exactly because
their Yukawa couplings to ordinary fermions are very small. The
latter fact is a key point for the baryogenesis in the $\nu$MSM
\cite{Akhmedov:1998qx,Asaka:2005pn}, ensuring the validity of the
third Sakharov condition. In \cite{Asaka:2005pn} was shown that the
$\nu$MSM can provide simultaneous solution to the problem of neutrino
oscillations, dark matter and baryon asymmetry of the Universe. 

{\em Inflation.} In \cite{Shaposhnikov:2006xi} it was proposed the
the $\nu$MSM may be extended by a {\em light} inflaton in order to
accommodate inflation. To reduce the number of parameters and to have
a common source for the Higgs and sterile neutrino masses the 
inflaton - $\nu$MSM couplings can be taken to be scale invariant on
the classical level. The mass of the inflaton can be as small as few
hundreds MeV, and the coupling of the lightest sterile neutrino to
the inflaton may serve as an efficient mechanism for the dark matter
production  for $M_1 < {\cal O}(10)$ MeV.

\begin{figure}
\centerline{\includegraphics[width=12cm]
{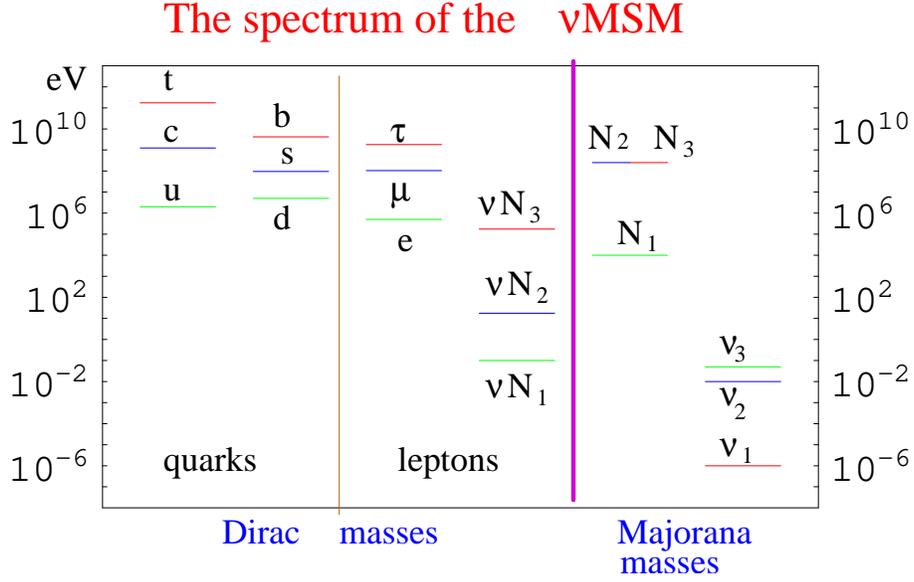}}
\caption{The expected mass spectrum of the $\nu$MSM. For quarks
and charged leptons the experimental data is used.}
\label{fig:spectrum}
\end{figure}

The pattern of the mass parameters, leading to successful
phenomenological predictions of the $\nu$MSM (neutrino masses and
oscillations, dark matter and baryon asymmetry of the Universe) is
shown schematically in Fig. \ref{fig:spectrum}.

\section{Search for DM sterile neutrino in the Universe}
\label{sec:univ} 
The dark matter sterile neutrino is likely to have a mass in the
${\cal O} (10)$ keV region. The arguments leading to the keV mass
scale are related  the problems of missing satellites and cuspy
profiles in the Cold Dark Matter cosmological models
\cite{Moore:1999nt,Bode:2000gq,Goerdt:2006rw,Gilmore:2007fy}; the keV
scale is also favoured by the cosmological considerations of the
production of dark matter due to transitions between active and
sterile neutrinos \cite{Dodelson:1993je,Shi:1998km}; warm DM may help
to solve  the problem of galactic angular
momentum~\cite{Sommer-Larsen:1999jx}. However, no upper limit on the
mass of sterile neutrino exists
\cite{Asaka:2006ek,Shaposhnikov:2006xi} as this particle can be
produced in interactions beyond the $\nu$MSM. An astrophysical lower
bound on their mass is $0.3$ keV, following from the analysis of the
rotational curves of dwarf spheroidal galaxies
\cite{Tremaine:1979we,Lin:1983vq,Dalcanton:2000hn}. Somewhat stronger
(but model dependent) lower bound on their mass can be derived from
the analysis of Ly-$\alpha$ clouds, limiting  their free streaming
length at the onset of cosmological structure formation
\cite{Hansen:2001zv,Viel:2005qj,Seljak:2006qw,Viel:2006kd}.  This
limit reads $M_1 >  M_{Ly\alpha}\left(\frac{<p_s>}{<p_a>}\right)$,
where $<p_s>$ and $<p_a>$ are the average momenta of the sterile and
active neutrinos correspondingly and  $M_{Ly\alpha} \simeq 10-14 $
keV \cite{Seljak:2006qw,Viel:2006kd}.

\begin{figure}
\hspace{2cm}\includegraphics[width=8cm]{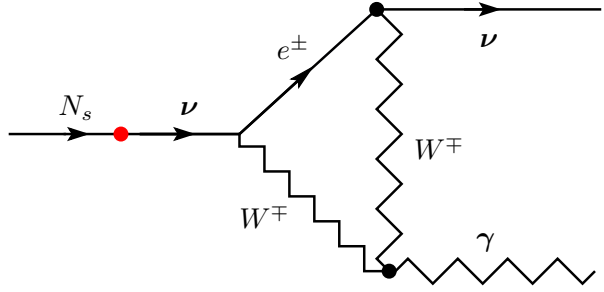}
\caption{The one-loop diagram giving rise to radiative neutrino decay.}
\label{fig:rad}
\end{figure}

In fact, the dark matter made of sterile neutrinos is not completely
dark, since there is a subdominant radiative decay channel (see Fig.
\ref{fig:rad}) $N_1\rightarrow \nu\gamma$  with the
width~\cite{Pal:1981rm,Barger:1995ty}
\be
 \Gamma_{\rm rad} = \frac{9\, \alpha_\textsc{em}\, G_F^2} {256\cdot
          4\pi^4}\:\sin^2(2\theta)\, M_1^5~,
\ee
where $G_F$ is the Fermi constant. These decays produce a narrow 
photon line (the width of it is determined by the Doppler effect due
to the motion of DM particles) with energy  $E_\gamma = \frac{M_1}{2}$.
This line  can be potentially observed in different X-ray
observations~\cite{Dolgov:2000ew,Abazajian:2001vt}.
Energy flux produced by the DM decay from a given direction into a
sufficiently narrow solid angle $\Omega \ll 1$ is given by
\begin{equation}
\label{eq:8} 
  F =  \frac{\Gamma_{\rm rad}\Omega}{8\pi}
  \hskip -3ex\int\limits_{{\rm line~ of~ 
      sight}} \hskip -3ex\rho_{DM}(r)dr \;,
\end{equation}
where $\rho_{DM}(r)$ is the DM density profile.

To optimize the search for DM sterile neutrino one should find the
astrophysical objects for which the value of integral (\ref{eq:8}) is
maximal whereas the  X-ray background is minimal. There is quite an
amazing empirical fact \cite{Boyarsky:2006fg} that  the signal  is
roughly the same  for many astrophysical objects,  from clusters to
dwarf galaxies. Namely, the  Milky Way halo signal is comparable with
that of clusters like Coma or Virgo, the  DM flux from Draco or Ursa
Minor dwarf spheroidals is 3 times stronger than that of the Milky
Way (MW) halo. At the same time, the  background strongly depends on
the astrophysical object. Indeed,  clusters of galaxies (e.g. Coma or
Virgo) have the temperature of intra-cluster media in keV range
leading  to strong X-ray emission contributing both to continuous
and  discrete (atomic lines) background spectrum. The continuum X-ray
emission from Milky Way is about two orders weaker than that of a
cluster, whereas dwarf satellites of the MW are really dark from the
point of view of X-ray background. Therefore, the best object to look
at to find the DM sterile neutrino are the  Milky Way and dwarf
satellite galaxies \cite{Boyarsky:2006fg}; X-ray quiet outer parts of
clusters can be used as well \cite{Boyarsky:2006zi}.

\begin{figure}
\centerline{\includegraphics[width=12cm]{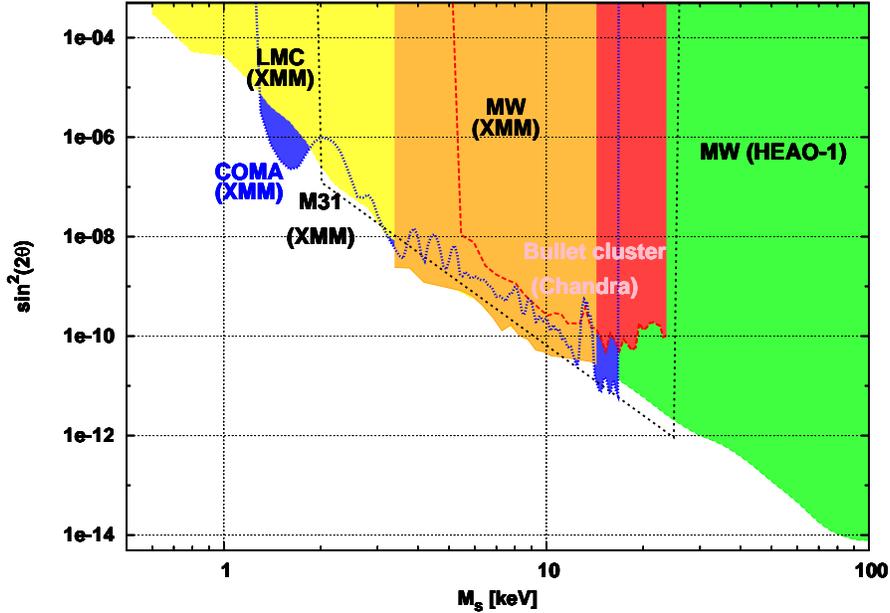}}
\caption{Upper bound on the mixing angle of dark matter sterile neutrino,
coming from X-ray observations of Large Magellanic Cloud (LMC) and
Milky Way (MW) by XMM-Newton and HEAO-1 satellites.}
\label{fig:bounds}
\end{figure}

Till now, no candidate for DM sterile neutrino decay line has been
seen and only the limits on the strength of their interaction with
active neutrinos exist. Over the last year restrictions on sterile
neutrino  parameters were improved by several orders of magnitude
\cite{Boyarsky:2005us,Boyarsky:2006zi,Boyarsky:2006fg,Riemer-Sorensen:2006fh,Watson:2006qb,Boyarsky:2006kc,Boyarsky:2006ag,Riemer-Sorensen:2006pi,Abazajian:2006jc,Boyarsky:2006hr}. 
The summary
of constraints on the mixing angle are shown on Fig.
\ref{fig:bounds}, where all results are subject to intrinsic factor 
$\sim 2$ uncertainty coming, in particular, from poor knowledge of
the dark matter distributions.

Unfortunately, the new data from {\em Chandra} and {\em XMM-Newton}
can hardly improve the constraints by more than a factor $10$ because
these instruments have the energy resolution exceeding greatly the
expected width of the DM line. To go much further one would need an 
improvement of spectral resolution up to the natural line width
($\Delta E/E \sim 10^{-3}$), have a reasonably wide field of view
$\sim 1^\circ$  (size of a dSph) and perform a wide energy scan, from
${\cal O}(100)$ eV to ${\cal O}(10)$ MeV. The discussion of
sensitivity of different existing and future Space missions can be
found in  \cite{Boyarsky:2006hr}.

\section{Search for DM sterile neutrino in laboratory
\cite{Bezrukov:2006cy}} 
\label{lab} 
Imagine now that some day an unidentified narrow line will be found
in X-ray observations.  Though there are a number of tests that could
help to distinguish the line coming from DM decays from the lines
associated with atomic transitions in interstellar medium, how can we
be sure that the dark matter particle is indeed discovered?  Clearly,
a laboratory experiment, if possible at all, would play a key role.
Current bounds in the interesting mass region were mostly based on
kink search in $\beta$-decay, inspired by possible discovery of 17~keV
neutrino.  The present bounds\cite{Yao:2006px} are much weaker
than required to compete with X-ray observations. 
Fig.~\ref{fig:bounds} demonstrates that the search for DM sterile
neutrino in terrestrial experiments is very challenging, as the
strength of interaction of DM sterile neutrino with the matter is
roughly $\theta^2$ times weaker than that of ordinary neutrino!

On very general grounds the possible experiments for the search of
sterile neutrinos can be divided in three groups:

{(i)} Sterile neutrinos are \emph{created} and subsequently
\emph{detected} in the laboratory.  The number of events that can be
associated with sterile neutrinos in this case is suppressed by
$\theta^4$ in comparison with similar processes with ordinary
neutrinos.  The smallness of the mixing angle, as required by X-ray
observations, makes this type of experiments hopeless.  For example,
for sterile neutrino mass $M_1=5~\keV$, the suppression in comparison
with neutrino reactions is at least of the order of $10^{-19}$.

(ii) Sterile neutrinos are created somewhere else in large amounts
and then \emph{detected} in the laboratory.  The X-ray Space
experiments are exactly of this type: the number density of sterile
neutrinos is fixed by the DM mass density, and the limits on the
X-ray flux give directly the limit on $\theta^2$ rather than
$\theta^4$ as in the previous case.  Another potential possibility is
to look for sterile neutrinos coming from the Sun.  The flux of
sterile neutrinos from, say, $pp$ reactions is
$F_N \sim 6\times 10^{10}~\theta^2/\mathrm{cm}^2\mathrm{s}$.  The only
way to distinguish sterile neutrinos from this source from electronic
neutrinos is the kinematics of the reactions $\nu_e n \rightarrow p
e$ and $N_1 n \rightarrow p e$, which looks hopeless.  For higher
energy sources, such as $^8B$ neutrinos, the emission of sterile $N_1$
would imitate the anti-neutrinos from the Sun due to the reaction $N_1 p
\rightarrow n e^+$ which is allowed since $N_1$ is a Majorana particle.
However, this process is contaminated by irremovable background from
atmospheric anti-neutrinos.  Even if all other sources of background
can be eliminated, an experiment like KamLAND would be able to place
a limit of the order of $\theta^4 < 3\times 10^{-7}$, which is weaker
than the X-ray limit for all possible sterile neutrino masses obeying
the Tremaine--Gunn bound.  The current KamLAND limit can be extracted
from \cite{Eguchi:2003gg} and reads $\theta^4 < 2.8\times 10^{-4}$. 
The sterile neutrino can also be emitted in supernovae (SNe)
explosions in amounts that could be potentially much larger than
$\theta^2 F_\nu$, where $F_\nu$ is the total number of active
neutrinos coming from SNe.  The reason is that the sterile neutrinos
interact much weaker than ordinary $\nu$ and thus can be emitted from
the volume of the star rather than from the neutrino-sphere.  Using
the results of \cite{Dolgov:2000ew}, the flux of SNe sterile
neutrinos due to $\nu_e-N$ mixing is $F_N \simeq
5 \times 10^{3}\theta^2\left(M_1/{\rm keV}\right)^4 F_\nu$.   In
spite of this enhancement, we do not see any experimental way to
distinguish the $N_1$ and $(\nu,~\bar\nu)$ induced events in the
laboratory.

(iii) The process of sterile neutrinos \emph{creation} is studied in
the laboratory.  In this case one can distinguish between two
possibilities.  In the first one, we have a reaction which would be
exactly forbidden if sterile neutrinos are absent.  We were able to
find just one process of this type, namely
$S\rightarrow{\rm invisible}$, where $S$ is any scalar boson.
Indeed, in the SM the process $S\rightarrow\nu\bar\nu$ is not allowed
due to chirality conservation, and $S\rightarrow\nu\nu$ is forbidden
by the lepton number conservation.  With sterile neutrinos, the
process $S\rightarrow \nu N_1$ may take place.  However, a simple
estimate shows that the branching ratios for these modes for
available scalar bosons such as $\pi^0$ or $K^0$ are incredibly small
for admitted (by X-ray constraints) mixing angles.  So, only one
option is left out: the detailed study of kinematics of different
$\beta$ decays.

An obvious possibility would be the main pion decay mode
$\pi \to \mu\nu$ with creation of sterile neutrino $N_1$ instead of the
active one.  This is a two body decay, so the energy muon spectrum is
a line with the kinetic energy $(m_\pi-m_\mu)^2/2m_\pi=4.1$ MeV for
decay with active neutrino and $((m_\pi-m_\mu)^2-M_1^2)/2m_\pi$ for
decay with massive sterile neutrino.  Thus, for $M_1$ of keV order
one needs the pion beam with energy spread less than 0.01~eV to
distinguish the line for sterile neutrino, which seems to be
impossible to get with current experimental techniques.

In the case of $\beta$-decay there are two distinct possibilities.  One
is to analyze the electron spectrum only.  In this case the admixture
of sterile neutrinos leads to the kink in the spectrum at the
distance $M_1$ from the endpoint.  However, the distinguishing a
small kink of the order of $\theta^2$ on top of the electron spectra
is very challenging  from the point of view of statistically large
physical background and nontrivial uncertainties in electron spectrum
calculations. The case of full kinematic reconstruction of $\beta$-decay
of radioactive nucleus is thus more promising.

The idea of using $\beta$-decay for sterile neutrino detection is quite
simple: measuring the full kinematic information for the initial
isotope, recoil ion, and electron one can calculate the neutrino
invariant mass on event by event basis.  In an ideal setup of exact
measurement of all these three momenta such an experiment provides a
background-free measurement where a single registered anomalous event
will lead to the positive discovery of DM sterile neutrino.  This
idea was already exploited at time of neutrino discovery and testing
of the Fermi theory of $\beta$ decay \cite{Pontecorvo1947}.  It was
also proposed to use full kinematic reconstruction to verify the
evidence for 17~keV neutrino found in the kink searches
\cite{Cook:1991cm,Finocchiaro:1992hy} to get rid of possible
systematics deforming the $\beta$-spectrum.  Recently, bounds on sterile
neutrino mixing were achieved by full kinematic reconstruction of
$^{38m}$K isotope confined in a magneto--optic trap
\cite{PhysRevLett.90.012501} but for a neutrino in the mass range
$0.7-3.5$~MeV, what is much heavier than considered here. For
$370-640$~keV mass range a similar measurement was performed in
electron capture decay of $^{37}$Ar \cite{Hindi:1998ym}.  We will
discuss below a possible setup for a dedicated experiment for a
search of keV scale DM sterile neutrino.

Let us consider an idealized experiment in which a cloud of
$\beta$-unstable nuclei, cooled to temperature $T$, is observed.  For
example, for $^3$H the normal $\beta$-decay is \be {}^3\mathrm{H} \to
{}^3\mathrm{He}+e+\bar{\nu}_e\;, \ee while in presence of sterile
neutrino in about $\theta^2$ part of the events (up to the kinematic
factor) the decay proceeds as \be {}^3\mathrm{H} \to
{}^3\mathrm{He}+e+N_1\;, \ee where $N_1$ is a sterile neutrino in
mostly right-handed helicity state. Suppose that it is possible to
register the recoil momentum of the daughter ion and of the electron
with high enough accuracy.  Indeed, existing  Cold-Target
Recoil-Ion-Momentum Spectroscopy (COLTRIMS) experiments are able to
measure very small ion recoil~\cite{Ullrich1997,Dorner2000}.  They
are utilized for investigation of the dynamics of ionization
transitions in atoms and molecules.  The ion momenta is determined by
time of flight measurement.  A small electric field is applied to the
decay region to extract charged ions into the drift region.  After
the drift region the ions are detected by a position sensitive
detector, which allows to determine both the direction of the momenta
and the time of flight.  Characteristic energies of recoil ion in
$\beta$-decay is of the order of the recoil momenta measured by
existing COLTRIMS in ion--atom collisions. Precisions currently
achieved with such apparatus are of the order of $0.2~\keV$ for the
ion momentum~\cite{Dorner2000,Dorner:97,Mergel:97,Dorner:98}.

Electron detection is more difficult, as far as the interesting
energy range is of the order of 10~keV for $^3$H decay (or greater
for most other isotopes).  This is much higher than typical energies
obtained in atomic studies.  One possible solution would be to use
the similar time-of-flight technique as for the recoil ions, but with
adding magnetic field parallel to the extraction electric field, thus
allowing to collect electrons from a wider polar angle.  In existing
applications such a method was used for electrons with energies of
only 0.1~keV \cite{Moshammer1996,Kollmus1997}.  In \cite{Jahnke2004}
retarding field was added in the electron drift region allowing to
work with electrons of up to 0.5~keV energies.  Alternatively, one
may try to use electrostatic spectrometers for electron energy
measurement, as it was proposed in
\cite{Cook:1991cm,Finocchiaro:1992hy}.  On the one hand, the latter
method allows to use the electron itself to detect the decay moment
for recoil time of flight measurement. On the other hand, it is hard
to reach high polar angle acceptance with this method, thus losing
statistics.

The decay moment needed for the time-of-flight measurement can be
tagged by registering the Lyman photon emission of the excited ion or
by the electron detection, if electron energy is determined by a
dedicated spectrometer.  Note, that for $^3\mathrm{H}_2$ case Lyman
photon is emitted only in about 25\% of the events
\cite{Finocchiaro:1992hy}, so the photon trigger also induces some
statistics loss.

According to \cite{Dorner2000} it is possible to achieve sensitivity
for measuring normal active neutrino masses of $10~\eV$ for each
single event; the accuracy needed for the case of sterile neutrinos
is considerably less than that as the mass of $N_1$ is expected to be
in the keV region.  Moreover, the measurement in the latter case is a
relative measurement, which is much simpler than absolute measurement
of the peak position required for active neutrino mass determination.

Of course, for a detailed feasibility study of $\beta$-decay
experiments to search for DM sterile neutrino a number of extra
points, including existence of possible backgrounds, must be
clarified. One obvious background appears from the fact that in
${\cal O}(\alpha_{\rm EM})$ cases of $\beta$-decays one gets an extra
photon, making the statistics requirement much stronger, unless  this
photon can be registered with 100\% efficiency. A very hard problem
is the low density of cold atoms (serving as a source of
beta-decays), available at present. Indeed, in order to compete with
X-ray mission in Space for $M_1 = 5$ KeV one should be able to
analyse the kinematics of about $10^{10}$ $\beta$-decays! Perhaps,
instead of tritium one can use other isotopes which have higher decay
energy release but are short lived, providing thus larger
statistics. 


\section{Search for singlet fermions in accelerator experiments
\cite{Gorbunov}} \label{accel}
In addition to DM sterile neutrino the $\nu$MSM contains a pair of
more heavier singlet fermions, $N_2$ and $N_3$. For an efficient
baryogenesis these particles must be almost degenerate in mass
\cite{Akhmedov:1998qx,Asaka:2005pn}. In addition, strong constraints
on the strength of interaction of these particles are coming from the
data on neutrino oscillations and
cosmology\cite{Gorbunov,Shaposhnikov:2006nn} (baryogenesis and Big
Bang Nucleosynthesis). In \cite{Shaposhnikov:2006nn} it was argued
that a specific mass-coupling pattern for the singlet fermions,
required for the phenomenological success of the $\nu$MSM, can be a
consequence of a lepton number symmetry, slightly broken by the
Majorana mass terms and Yukawa coupling constants. The existence of
this symmetry provides an argument in favour of ${\cal O} (1)$ GeV
mass of these neutral leptons and makes the couplings $F_{\alpha2}$
of singlet fermions to ordinary leptons considerably enhanced in
comparison with a naive estimate $F^2\sim M_2 \sqrt{\Delta m^2_{\rm
atm}}/v^2$. It is interesting to know, therefore,
what would be the experimental signatures of the neutral singlet
fermions in this mass range and in what kind of experiments they
could be found.

Naturally, several distinct strategies can be used for the
experimental search of these particles. The first one is related to
their production. The singlet fermions  participate in all reactions
the ordinary neutrinos do with a probability suppressed roughly by a
factor $(M_D/M_M)^2$, where $M_D$ and $M_M$ are the Dirac and
Majorana masses correspondingly. Since they are massive, the
kinematics of, say, two body decays $K^\pm \rightarrow \mu^\pm N$,
$K^\pm \rightarrow e^\pm N$ or three-body decays  $K_{L,S}\rightarrow
\pi^\pm + e^\mp + N_{2,3}$ changes  when $N_{2,3}$ is replaced by an ordinary
neutrino. Therefore, the study of  {\em kinematics} of rare meson
decays can constrain the strength of the coupling of heavy leptons.
This strategy has been used in a number of experiments for the search
of neutral leptons in the past \cite{Yamazaki:1984sj,Daum:2000ac},
where the spectrum of electrons or muons originating in decays $\pi$
and $K$ mesons has been studied. The second strategy is to look for
the  decays of neutral leptons inside a detector 
\cite{Bernardi:1985ny,Bernardi:1987ek,Vaitaitis:1999wq,Astier:2001ck}
(``nothing" $\rightarrow$ leptons and hadrons).  Finally, these two
strategies can be unified, so that the production and the decay
occurs inside the same detector \cite{Achard:2001qw}.

Clearly, to find the best way to search for neutral leptons, their
decay modes have to be identified and branching ratios must be
estimated. A lot of work in this direction has been already done in
Refs. 
\cite{Shrock:1980ct,Shrock:1981wq,Gronau:1984ct,Johnson:1997cj} for
the general case, ref. \cite{Gorbunov}  deals with a specific case of
the $\nu$MSM.

We arrived to the following conclusions~\cite{Gorbunov}.\\ 

(i) The singlet fermions with the masses smaller than $m_\pi$ are
already disfavoured on the basis of existing experimental data of 
\cite{Bernardi:1987ek} and from the requirement that these particles
do not spoil the predictions of the Big Bang Nucleosynthesis (BBN)
\cite{Dolgov:2000jw,Dolgov:2000pj} (s.f. \cite{Kusenko:2004qc}). 

\begin{figure}
\centerline{
\includegraphics[width=5cm]{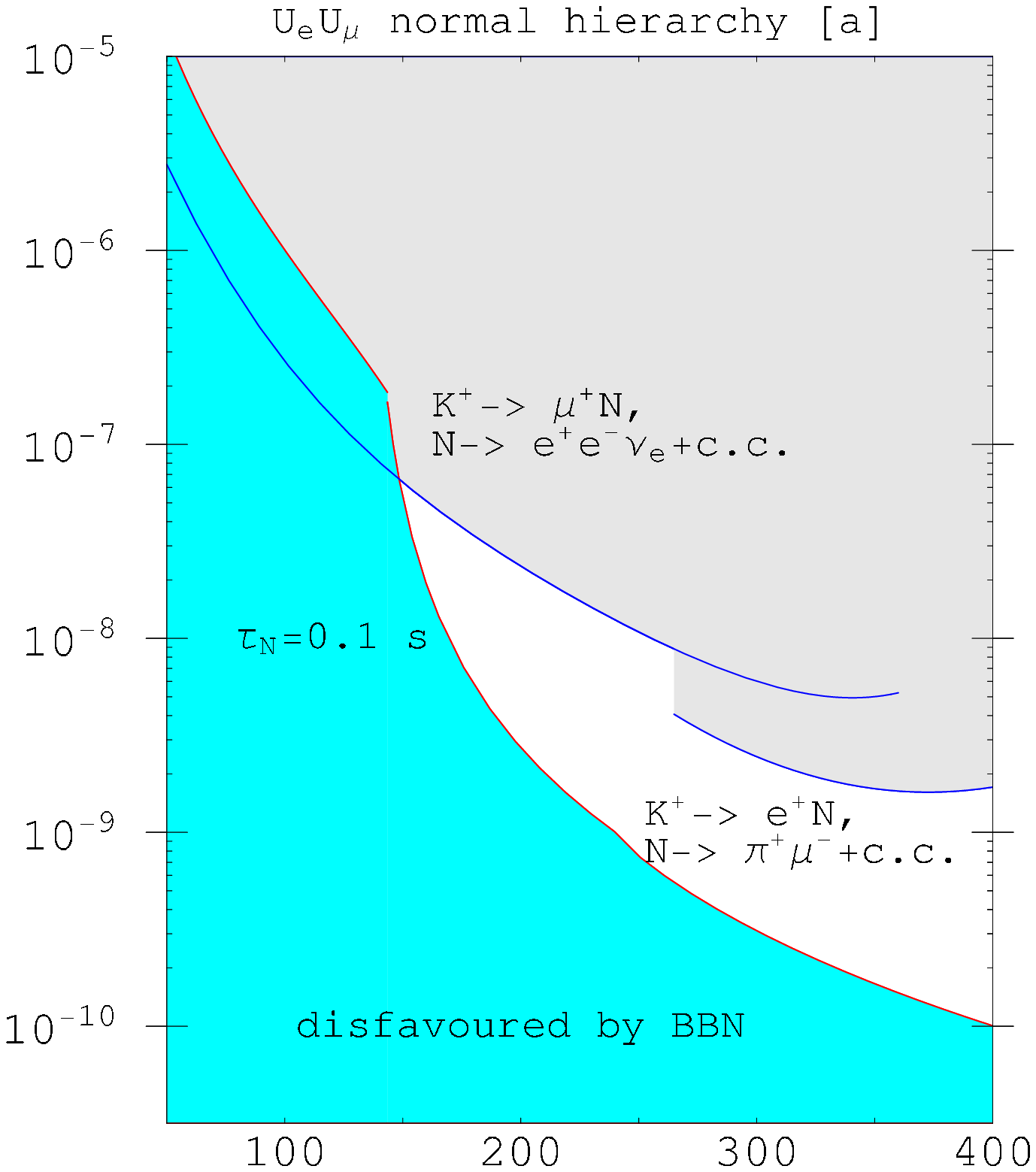}
\includegraphics[width=5cm]{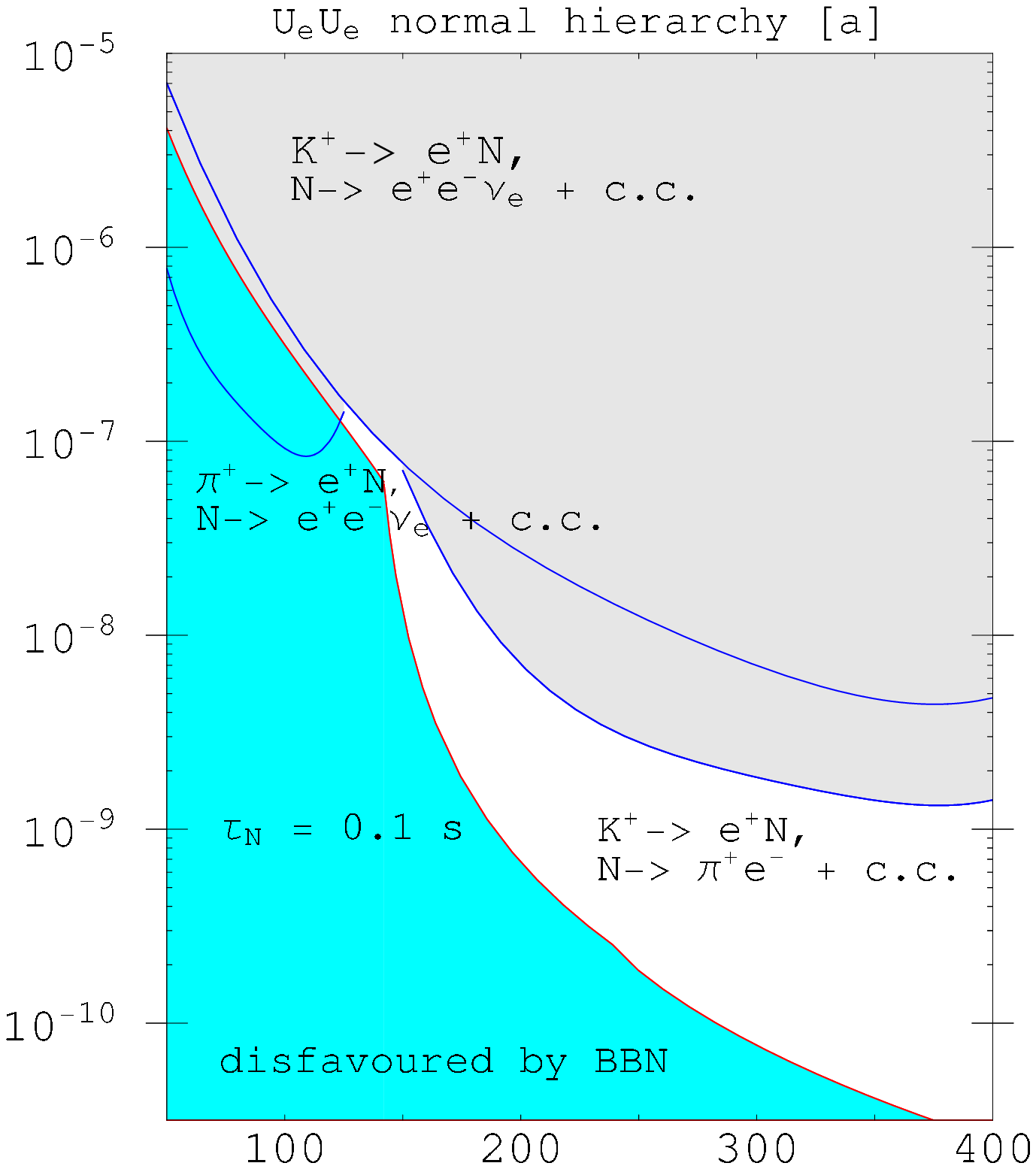}
\includegraphics[width=5cm]{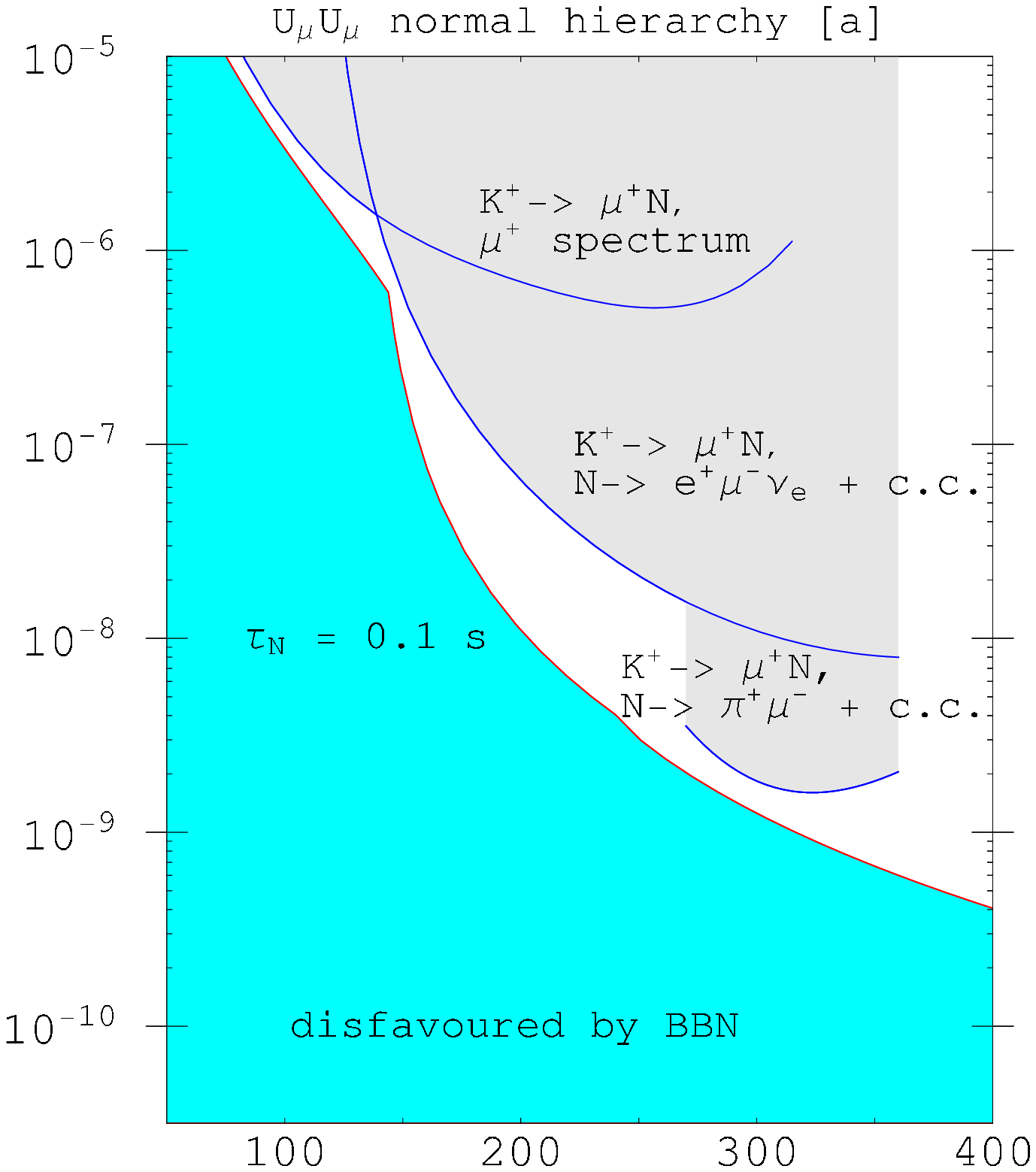}}
\caption{Experimental (upper bound) and BBN (lower bound) constraints
on the mass and mixing angles $U_\alpha=M_D^{\alpha2}/M_2$ of the
singlet fermions with masses below $400$ MeV in the $\nu$MSM.  Blank
regions  are phenomenologically allowed. }
\label{fig:heavy}
\end{figure}

(ii) The mass interval $m_\pi < M_N < m_K$ is perfectly allowed from
the cosmological and experimental points of view, see Fig.
\ref{fig:heavy}.
Moreover, further constraints on the couplings of singlet fermions
can be derived from the reanalysis of the {\em already existing but
never considered from this point of view} experimental data of KLOE
collaboration and of the E787 experiment. In addition, the NA48/3
(P326) experiment at CERN would allow to find or to exclude
completely singlet fermions with the mass below that of the kaon. The
search for the missing energy signal, specific for the experiments
mentioned above, can be complimented by the search of decays of
neutral fermions, as was done in CERN PS191 experiment
\cite{Bernardi:1985ny,Bernardi:1987ek}. To this end quite a number of
already existing or planned neutrino facilities (related, e.g. to
CNGS, MiniBoone, MINOS or J-PARC), complemented by a near dedicated
detector (like the one of CERN PS191) can be used.

(iii) In the mass interval $m_K < M_N < 1$ GeV the detailed study of
kinematics of decays of charmed mesons and $\tau$ leptons, possible
at charm and $\tau$ factories, can enter into cosmologically
interesting part of the parameter space of the $\nu$MSM.

(iv) For $1$ GeV $< M_N < m_D$ the search for the specific missing
energy signal, potentially possible at beauty, charm and $\tau$
factories, is unlikely to gain the necessary statistics and is very
difficult if not impossible at hadronic machines like LHC. So, the
search for decays of neutral fermions is the most effective
opportunity. In short, an intensive beam of protons, hitting the
fixed target, creates, depending on its energy, pions, strange and
charmed  mesons that decay and produce heavy neutral leptons. A part
of these leptons then decay in a detector, situated some distance
away from the collision point. The  dedicated experiments on the
basis of the proton beam NuMI or NuTeV at FNAL, CNGS at CERN, or
J-PARC  can touch a very interesting parameter range for $M_N \lsim
1.8$ GeV.

(v) Going above $D$-meson but still below $B$-meson thresholds is
very hard if not impossible with present or planned proton machines or
B-factories. To enter into cosmologically interesting parameter space
would require the increase of the present intensity of, say, CNGS
beam by two orders of magnitude or to producing and studying the
kinematics of more than $10^{10}$ B-mesons.

\section{Conclusions}

To conclude, none of the {\em experimental observations}, which are
sometimes invoked as the arguments for the existence of the large
$\sim 10^{10}-10^{15}$ GeV intermediate energy scale between the
$W$-boson mass and the Planck mass, really requires it. The smallness
of the active neutrino masses may find its explanation in small
Yukawa couplings rather than in large energy scale. The dark matter
particle, associated usually with a WIMP of  ${\cal O}(100)$ GeV mass
or an axion, can well be a much lighter sterile neutrino, practically
stable on the cosmological scales. The thermal leptogenesis
\cite{Fukugita:1986hr}, working well only at large masses of Majorana
fermions, can be replaced by the baryogenesis through light singlet
fermion oscillations. The inflation can be associated with the light
inflaton field rather than with that with the mass $\sim 10^{13}$
GeV, with the perturbation power spectrum coming from inflaton
self-coupling rather than from its mass.

Inspite of the fact that all new particles of the $\nu$MSM are light,
it is a challenge to uncover them experimentally due to the extreme
weakness of their interactions. To search for dark matter sterile
neutrinos in the Universe one needs an X-ray spectrometer in Space 
with good energy resolution $\delta E/E \sim 10^{-3}-10^{-4}$ getting
signals from our Galaxy and its dwarf satellites. The laboratory
search for these particles would require the detailed analysis of
kinematics of $\beta$-decays of different isotopes, which is
extremely hard. The search for heavier singlet fermions, responsible
for baryon asymmetry of the Universe is relatively easy if they are
lighter than K-meson, possible with existing accelerators if the are
lighter than D-mesons, and extremely challenging if they have a mass
above $2$ GeV.

At the same time, the $\nu$MSM can be falsified by a number of
different experiments. For example, the discovery of WIMPs in dark
matter searches, supersymmetry or any new particle except the Higgs
boson at LHC, confirmation of the LSND result (according to MiniBooNE
publication \cite{Aguilar-Arevalo:2007it} that appeared after this
conference the neutrino oscillation explanation of the LSND anomaly
is rejected at the $98$\% confidence level) or confirmation of the
claim \cite{Klapdor-Kleingrothaus:2006ff} on the observation of
neutrino-less double $\beta$-decay (this process was considered in
the framework of the $\nu$MSM in \cite{Bezrukov:2005mx}) would
disprove the $\nu$MSM. The same conclusion is true if the active
neutrinos are found to be degenerate in mass.

\section{Acknowledgements} This work was supported in part by the
Swiss National Science Foundation. I thank Takehiko Asaka, Steve
Blanchet, Fedor Bezrukov, Alexey Boyarsky, Dmitry Gorbunov, Alexander
Kusenko, Mikko Laine, Andrei Neronov, Oleg Ruchayskiy and Igor
Tkachev for collaboration.



\end{document}